\documentclass[10pt]{article}
\usepackage[utf8]{inputenc}
\usepackage[T1]{fontenc}
\usepackage[top=1in,bottom=1in,left=1in,right=1in]{geometry}
\usepackage{authblk}
\usepackage{amsmath,amsfonts,amssymb,amsthm}

\numberwithin{equation}{section}

\newtheorem{theorem}{Theorem}
\newtheorem{proposition}{Proposition}

\numberwithin{equation}{section}

\newcommand{\C}{\mathbb{C}}

\newcommand{\ket}[1]{\rvert#1\rangle}
\newcommand{\bra}[1]{\langle #1\rvert}
\newcommand{\braket}[2]{\langle #1\rvert#2\rangle}

\newcommand{\qbin}[3]{\begin{bmatrix} #1 \\ #2 \end{bmatrix}_{#3}}
\newcommand{\qpoc}[3]{\left(#1;#2\right)_{#3}}
\newcommand{\pfq}[4]{\left(\begin{matrix} #1 \\ #2 \end{matrix};#3,#4 \right)}

\newcommand{\qg}{SU_{q}(3)}

\newcommand{\qm}{q^{-2}}
\newcommand{\qp}{q^{2}}

\title{$SU_q(3)$ corepresentations and bivariate q-Krawtchouk polynomials.}
\author[1]{Geoffroy Bergeron}
\author[2]{Erik Koelink}
\author[1]{Luc Vinet}
\affil[1]{Centre de recherches math\'ematiques, Universit\'e de Montr\'eal, P.O. Box 6128, Centre-ville Station, Montr\'eal, Canada H3C 3J7}
\affil[2]{IMAPP, Radboud Universiteit, P.O. Box 9010, 6500 GL Nijmegen, The Netherlands}

\date{\today}
\begin{document}
\maketitle
\thispagestyle{empty}
\hrule
\begin{abstract}\noindent
The matrix elements of unitary $SU_q(3)$ corepresentations, which are analogues of the symmetric powers of the natural repesentation, are shown to be the bivariate $q$-Krawtchouk orthogonal polynomials, thus providing an algebraic interpretation of these polynomials in terms of quantum groups.
\end{abstract} 
\hrule

\section*{Introduction}
A fruitful connection exists between Lie groups and algebras and the theory of orthogonal polynomials. Algebraic interpretations for these orthogonal polynomials enable simple derivations of their properties and often lead to new identities. Similar connections between the theory of quantum groups and (mostly univariate) $q$-orthogonal polynomials have been established \cite{Koornwinder1990}. The results of this paper pursue such a connection in multivariate situations by giving an algebraic interpretations of the quantum bivariate $q$-Krawtchouk polynomials in terms of the quantum group $SU_q(3)$.

In the classical case, the Krawtchouk polynomials of the Tratnik type form a family of multivariate Krawtchouk polynomials constructed from univariate Krawtchouk polynomials using a construction developed in \cite{Tratnik1991} that applies to all polynomials of the ($q=1$) Askey scheme. These are orthogonal polynomials with respect to the multinomial distribution. Many Lie-theoretic interpretations have been given for the Krawtchouk polynomials. The multivariate Krawtchouk polynomials in $d$ variables were shown \cite{Genest2013} to be matrix elements of the $SO(d+1)$ Lie group and identified as well \cite{Iliev2012a,Iliev2012} as overlaps of anti-automorphisms of $\mathfrak{sl}(d+1)$-modules.

In the context of quantum groups and algebras, interpretations analogous to the classical ones have been given for the $q$-Krawtchouk polynomials, which are orthogonal with respect to the $q$-binomial distribution. Koornwinder obtained \cite{Koornwinder1989} the univariate $q$-Krawtchouk polynomials as the matrix elements of unitary corepresentations of the $SU_q(2)$ quantum group. In a complementary way based on the quantum algebras, the $q$-Krawtchouk polynomials were seen \cite{Zhedanov1993,Genest2016} to arise as matrix elements of a class of $U_q(\mathfrak{sl}(2))$ automorphisms. These two approaches are essentially dual one to another \cite{Floreanini1993a}.

A family of multivariate $q$-Krawtchouk polynomials were first derived by Gasper and Rahman in \cite{Gasper2007} where they constructed $q$-deformations of Tratnik's polynomials. We will thus refer to these multivariate extensions of the $q$-Krawtchouk as being of the Tratnik type. An interpretation of the bivariate and multivariate $q$-Krawtchouk polynomials based on the quantum algebra viewpoint was obtained in \cite{Genest2017}. With an eye to generalizations and in view of the fact that for $q=1$, Lie groups rather than algebras provide a most natural framework, it seems appropriate to examine how the multivariate $q$-Krawtchouk polynomials can be obtained and analyzed in a quantum group framework. In this paper, we build upon the quantum group approach of Koornwinder to obtain the bivariate quantum $q$-Krawtchouk polynomials of the Tratnik type as the matrix element of unitary $SU_q(3)$ corepresentations. Within this quantum group approach, the structure of the unitary elements of $U_q(\mathfrak{sl}(3))$ constructed in \cite{Genest2017} is explained from the representation theory of $SU_q(3)$.

This paper is organized as follows. In section \ref{reps}, a presentation of the $\qg$ algebra is first given and the construction of its unitary representations is reviewed. Symmetric $SU_q(3)$ corepresentations are then constructed at the beginning of section \ref{corep}, followed by the derivation of their matrix elements and a proof of the unitarity of the corepresentations. A generating function for the matrix elements is then obtained. In section \ref{mxinreps}, the matrix elements are evaluated in irreducible $SU_q(3)$ representations and identified as bivariate $q$-Krawtchouk polynomials, which follows from Soibelman's tensor product theorem. Finally, section \ref{redrep} illustrates how evaluating the matrix elements in reducible $SU_q(3)$ representations leads to identities for orthogonal polynomials. This is followed by a brief conclusion.

\section{The $SU_q(3)$ Hopf algebra and its representations}\label{reps}
We first give in this section a presentation of the $\qg$ quantum group, a Hopf $*$-algebra, and discuss how its representations are constructed.

\subsection{The coordinate ring $A(SU_{q}(3))$}
The coordinate ring $A(SL_q(3;\mathbb{C}))$ is a $\mathbb{C}$-algebra $A=\mathbb{C}[x_{ij};1\leq i,j\leq3]$ with the relations
\begin{align}\label{suq3relations}
 x_{ik}x_{jk}&=q x_{jk}x_{ik}, \quad &x_{ki}x_{kj}&=q x_{kj}x_{ki}, \quad &\forall\quad i < j, \\
 x_{il}x_{jk}&=x_{jk}x_{il}, \quad &x_{ik}x_{jl}-q x_{il}x_{jk}&=x_{jl}x_{ik}-\frac{1}{q}x_{jk}x_{il}, \quad &\forall\quad i < j, \quad k < l, \nonumber
\end{align}
\begin{align*}
 \sum\limits_{\sigma \in S_3} (-q)^{l(\sigma)}x_{1\sigma(1)}x_{2\sigma(2)}x_{3\sigma(3)} = 1.
\end{align*}
A Hopf algebra structure is given with the following coproduct $\Delta$, counit $\epsilon$ and antipode $S$ 
\begin{align}\label{hopfSUq3}
 \Delta(x_{ij})&= \sum\limits_{k=1}^{3}x_{ik}\otimes x_{kj}, \quad \epsilon(x_{ij})=\delta_{ij}, \quad S(x_{ij}) = (-q)^{i-j} \xi_{ji}, \quad 1 \leq i,j \leq 3,
\end{align}
where $\xi_{ij}$ denotes the $(i,j)$ quantum minor, that is the quantum determinant of $x$ with the $i^{th}$ row and the $j^{th}$ column removed:
$$ \xi_{ij} = \sum\limits_{\tau \in S_{2}} (-q)^{l(\tau)} x_{i_{1} j_{\tau(1)}} x_{i_{2} j_{\tau(2)}}, \quad i_{1} < i_2 \in \{ 1,2,3 \} \setminus \{ i \}, \quad j_{1} < j_2 \in \{ 1,2,3 \} \setminus  \{ j \}.$$
Morevover, a unique conjugate linear anti-homomorphism $* :A(SL_q(3;\mathbb{C})) \rightarrow A(SL_q(3;\mathbb{C})) : x\mapsto x^{\ast}$ exists such that
\begin{align}\label{starop}
  x_{ij}^\ast = S(x_{ji}) = (-q)^{j-i} \xi_{ij}, \quad \forall \quad i,j.
\end{align}
This $\ast$-operation makes $A(SL_q(3))$ into the $\ast$-Hopf algebra $A(SU_q(3))$ which we will refer to as simply $SU_q(3)$.

\subsection{$\qg$ representations}
The $\qg$ $*$-representations used in this paper were constructed in \cite{Bragiel1989}. However, to obtain our main results, we will make use of a theorem of Soibelman \cite{Soibelman1990} on the construction of modules over quantum groups. Thus, we review how $*$-representations of unitary quantum groups are constructed \cite{Koelink1991} using this result. Following \cite{Korogodski1998}, one first defines \cite{Vaksman1988} the infinite dimensional representations $\tau_{\alpha}$ with $\alpha \in U(1)$, of $SU_q(2)$ \cite{Woronowicz1987} where its generators $\{ t_{ij}\, : \, i,j =1,2 \}$ act on a Hilbert space $H$ with orthonormal basis $\{\ket{n}, n \in \mathbb{N}\}$ as follows
\begin{align*}
 \tau_{\alpha}(t_{12})\ket{n} &= - q^{n+1} \alpha^{-1} \ket{n}, & \tau_{\alpha}(t_{21}) \ket{n} &= q^n \alpha \ket{n},\\
 \tau_{\alpha}(t_{11})\ket{n} &= \sqrt{1-q^{2n}}\ket{n-1},\quad \tau_{\alpha}(t_{11})\ket{0}=0, & \tau_{\alpha}(t_{22}) \ket{n} &= \sqrt{1-q^{2(n+1)}}\ket{n+1}.
\end{align*}
From these representations, one can build $\qg$ representations. Indeed, consider the two canonical embeddings $\varphi_i:U_q(\mathfrak{sl}(2))\hookrightarrow U_q(\mathfrak{sl}(3))$, $i =1,2$. These embeddings define by duality the projections $\varphi^*_i:\qg \longrightarrow SU_q(2)$ such that irreducible $*$-representations of $\qg$ are given by the maps $\pi_i \equiv \tau_{\alpha} \circ \varphi^*_i$ with $\alpha=-1$ acting on $V_{s_i}\cong H$. Explicitly, these elementary representations are specified by
\begin{align}\label{pi1rep}
 \pi_{1}
 \begin{pmatrix}
  x_{11} & x_{12} & x_{13}\\
  x_{21} & x_{22} & x_{23}\\
  x_{31} & x_{32} & x_{33}
 \end{pmatrix}
\cdot\,\ket{k}=
\left(
\begin{array}{ccc}
 \sqrt{1-q^{2 k}}|k-1\rangle & q^{k+1}\ket{k} & 0 \\
 - q^k |k\rangle & \sqrt{1-q^{2 k+2}}|k+1\rangle & 0 \\
 0 & 0 & |k\rangle  \\
\end{array}
\right),
\end{align}
and
\begin{align}\label{pi2rep}
 \pi_{2}
 \begin{pmatrix}
  x_{11} & x_{12} & x_{13}\\
  x_{21} & x_{22} & x_{23}\\
  x_{31} & x_{32} & x_{33}
 \end{pmatrix}
\cdot\,\ket{k}=
\left(
\begin{array}{ccc}
 |k\rangle  & 0 & 0 \\
 0 & \sqrt{1-q^{2 k}}|k-1\rangle & q^{k+1}\ket{k} \\
 0 & - q^{k} |k\rangle & \sqrt{1-q^{2 k+2}}\ket{k+1} \\
\end{array}
\right).
\end{align}
All $\qg$ representations can be constructed from the elementary representations $\pi_1$ and $\pi_2$ using the following tensor product theorem \cite[Thm 6.2.7]{Korogodski1998}. Denoting by $\C[G]_q$ the quantised algebra of functions \cite{Chari1994} on $G$, a connected and simply connected simple compact Lie group with associated Weyl group $W$, one has
\begin{theorem}[Tensor product theorem]\label{tensorprod}
 For any unitarizable irreducible $\C[G]_q$ $*$-representation $V$, there exists a unique element $w\in W$ of the Weyl group and a unique element $\tau \in T$ of the distinguished maximal torus such that
\begin{align*}
  V \cong V_w \otimes V_\tau, \qquad  V_w = V_{s_{i_1}}\otimes V_{s_{i_2}}\otimes \dots \otimes V_{s_{i_k}},
\end{align*}
 where $w=s_{i_1} s_{i_2} \dots s_{i_k}$ is a reduced decomposition of $w$. The tensor product does not depend (up to a unitary equivalence) on the choice of reduced decomposition for $w$.
\end{theorem}
The map $\pi_w:\qg \rightarrow End(V_w)$ associated to the representation in the above theorem is specified by
\begin{align*}
 \pi_w = (\pi_{i_1}\otimes\pi_{i_2}\otimes\dots\otimes\pi_{i_k})\circ \Delta^{(k-1)}, \qquad w=s_{i_1} s_{i_2} \dots s_{i_k},
\end{align*}
where the repeated coproduct $\Delta^{(k-1)}$ is defined through
\begin{align*}
\Delta^{(1)} = \Delta, \quad \Delta^{(k)} = \underbrace{1\otimes \dots \otimes 1}_{k-1\text{ times}} \otimes \Delta \circ \Delta^{(k-1)}.
\end{align*}
The representation on $V_\tau$ is a one-dimensional representation of the form $\rho_\tau:SU_q(3)\rightarrow \C:x_{ij} \mapsto \alpha_i(\tau) \delta_{ij}$ with $\alpha_i(\tau) \in U(1)$ such that $\alpha_1(\tau)\alpha_2(\tau)\alpha_3(\tau)=1$. Thus, it only contributes a global phase to our result and will not be further considered. Take now the case where $w=s_2 s_1$. One has
\begin{align}\label{pi12}
 \pi_{21} \equiv \pi_{s_2 s_1} = (\pi_2 \otimes \pi_1) \circ \Delta. 
\end{align}
The explicit action on the generators is easily obtained from \eqref{pi12} using \eqref{hopfSUq3}, \eqref{pi1rep} and \eqref{pi2rep}.

The representation $\pi_{121}$ is obtained similarly and, upon tensoring with one-dimensional representations, is the most general representation in the sense that the intersection of the kernels of these representation is trivial, c.f. e.g. \cite[Sect. 5]{Koelink1991}. It follows from Theorem \ref{tensorprod} that $\pi_{121}$ and $\pi_{212}$ are equivalent, so it suffices to consider one of them.

\section{Unitary $\qg$ corepresentations}\label{corep}
We now turn to the construction of unitary $\qg$ corepresentations in analogy with the $GL(3)$ coaction on functions on $\C^3$. Consider the space $\mathcal{F}^{(1)}(\C^3)$ of linear functions on $\C^3$ with orthonormal basis $\{ z_i \}_{i=1,2,3}$. By identifying these basis elements with a fixed column of the $\qg$ quantum group as follows
\begin{align}\label{zidentification}
 z_i = x_{ij},\qquad \text{for }j\text{ fixed,}
\end{align}
a natural \cite{Noumi1993} coaction is defined using the coproduct \eqref{hopfSUq3} as
\begin{align}\label{lincoaction}
 \Delta:\mathcal{F}^{(1)}(\C^3) \rightarrow \qg \otimes \mathcal{F}^{(1)}(\C^3) : z_i \mapsto \Delta(x_{ij}) = \sum\limits_{k=1}^3 x_{ik}\otimes x_{kj} \equiv \sum\limits_{j=1}^3 x_{ik}\otimes z_k.
\end{align}
The algebra $\mathcal{F}(\C^3)$ of polynomial functions on $\C^3$ is identified with the tensor algebra of $\mathcal{F}^{(1)}(\C^3)$ as follows
\begin{align}\label{tensoralgebra}
 \mathcal{F}(\C^3) = T(\mathcal{F}^{(1)}(\C^3)), \quad T(V) \equiv \bigoplus\limits_{n=0}^{\infty} V^{\otimes n}, \quad V^{\otimes n} \equiv \underbrace{V\otimes\dots\otimes V}_{n \text{ times}}, \quad V^{\otimes 0}\cong\C.
\end{align}
The identification \eqref{zidentification} together with \eqref{suq3relations} establishes the following relations
\begin{align}\label{qrelations}
 \mathcal{R} = \{\, z_i z_j \sim q z_j z_i \,\mid \, i < j,\,\, i,j = 1,2,3 \,\}.
\end{align}
Denote the quotient \cite[Ch. 7]{Chari1994} of the tensor algebra \eqref{tensoralgebra} by the relations \eqref{qrelations} as $\mathrm{Sym}_q(\C^3)$. A natural grading on $\mathrm{Sym}_q(\C^3)$ is inherited from the one of $\mathcal{F}(\C^3)$ as the relations \eqref{qrelations} preserve this grading. Explicitly,
\begin{align}\label{grading}
 \mathrm{Sym}_q(\C^3) = \bigoplus\limits_{n=0}^{\infty} \mathcal{F}_q^{(n)}(\C^3), \qquad \mathcal{F}_q^{(n)}(\C^3)\equiv \left( \mathcal{F}^{(1)}(\C^3) \right)^{\otimes n} \Big/ \mathcal{R}\,.
\end{align}
The coproduct being an homomorphism, the coaction \eqref{lincoaction} is extended to $\mathrm{Sym}_q(\C^3)$ as follows
\begin{align}\label{coaction}
 \Delta(z_1^{m_1}z_{2}^{m_2}z_3^{m_3}) = \Delta(z_1)^{m_1} \Delta(z_2)^{m_2} \Delta(z_3)^{m_3}, \qquad \forall\, z_1^{m_1}z_{2}^{m_2}z_3^{m_3} \in  Sym_q(\C^3).
\end{align}
Being constructed from the coproduct, \eqref{coaction} defines an $\qg$ corepresentation. From \eqref{lincoaction}, it is easily seen that the coaction \eqref{coaction} preserves the natural grading \eqref{grading} on $\mathrm{Sym}_q(\C^3)$. Thus, $\mathrm{Sym}_q(\C^3)$ as an $\qg$ corepresentation decomposes as a direct sum of corepresentations as follows
\begin{align*}
 \Delta:\mathrm{Sym}_q(\C^3)\longrightarrow \bigoplus\limits_{n=0}^{\infty} \Delta\left(\mathcal{F}_q^{(n)}(\C^3) \right).
\end{align*}
In view of \eqref{qrelations}, a basis for $\mathcal{F}_q^{(N)}(\C^3)$ is given by
\begin{align}\label{monomialbasis}
 \mathcal{B}_N = \left\{ \, z^{\vec{m}}\, \big{|} \, \rvert \vec{m} \rvert = N\, \right\}, \quad\text{for}\quad z^{\vec{m}}\equiv z_1^{m_1}z_{2}^{m_2}z_3^{m_3} \quad\text{and}\quad \rvert\vec{m}\rvert \equiv m_1+m_2+m_3.
\end{align}

\subsection{Matrix elements}
On the basis \eqref{monomialbasis}, the matrix elements of the corepresentation $\mathcal{F}_q^{(N)}(\C^3)$ are given in the following proposition.
%%%%%%%%%%%%%%%%%%%%%%%%%%%%%%%%%%%%%%%%%%%%%%%%%
\begin{proposition}[Matrix elements]\label{mxelements}
Let $0 < N \in \mathbb{N}$. The matrix elements $h_{\vec{m},\vec{n}}^{(N)}$ of the $\qg$ corepresentation $\mathcal{F}_q^{(N)}(\C^3)$ with coaction $\Delta$ are given in the basis $\mathcal{B}_N$ by
\begin{align}\label{unormalizedMxLeft}
\Delta(z^{\vec{m}})=\sum\limits_{\rvert\vec{n}\rvert=N} h_{\vec{m},\vec{n}}^{(N)}\otimes z^{\vec{n}}, \qquad h_{\vec{m},\vec{n}}^{(N)} = \sum\limits_{\substack{\,\rvert \overline{a}_j \rvert = m_j\\ \rvert \underline{a}_j \rvert = n_j}} Q(a) \qbin{\vec{m}}{a}{\qm} \prod\limits_{k=1}^{3}\left( \prod\limits_{i=1}^{3} x_{ik}^{a_{ik}} \right),
\end{align}
with $\rvert\overline{a}_i\rvert \equiv \sum_{j=1}^{3} a_{ij}$, $\,\rvert \underline{a}_j \rvert \equiv \sum_{i=1}^{3} a_{ji}$ and
\begin{align*}
 \qbin{\vec{m}}{a}{q} &\equiv \qbin{m_1}{\overline{a}_1}{q}\qbin{m_2}{\overline{a}_2}{q}\qbin{m_3}{\overline{a}_3}{q}, \qquad \qbin{m}{\overline{a}_i}{q} \equiv \frac{\qpoc{q}{q}{m}}{\qpoc{q}{q}{a_{i1}}\qpoc{q}{q}{a_{i2}}\qpoc{q}{q}{a_{i3}}},
\end{align*}
and where
\begin{multline}\label{Qa}
Q(a) = q^{-f(a)}, \quad f(a) = a_{13}(a_{21}+a_{22}+a_{32})+a_{31}(a_{12}+a_{22}+a_{23})\\+a_{12}a_{21}+a_{13}a_{31}+a_{23}a_{32}
\end{multline}
is a manifestly symmetric function of the matrix of indices.
\begin{proof}
First, compute the matrix elements of the coaction on powers of a single generator of $\mathcal{F}^{(1)}(\C^3)$. From \eqref{lincoaction} and \eqref{coaction}, one has
\begin{align*}
 \Delta(z_i^{m}) = \Delta(x_{ij})^{m} = \left(\sum\limits_{k=1}^{3} x_{ik}\otimes x_{kj}\right)^{m},
\end{align*}
and, knowing that $(x_{ik}\otimes x_{kj})(x_{il}\otimes x_{lj})= q^{2} (x_{il}\otimes x_{lj})(x_{ik}\otimes x_{kj})$, $\forall\, k<l$, one has that \cite{Koornwinder1989}
\begin{align*}
 \Delta(z_{i}^{m})= \sum\limits_{\rvert\overline{a}_i\rvert=m} \qbin{m}{\overline{a}_i}{\qm} x_{i1}^{a_{i1}}x_{i2}^{a_{i2}}x_{i3}^{a_{i3}} \otimes z_1^{a_{i1}} z_2^{a_{i2}} z_3^{a_{i3}},
 %= \sum\limits_{\rvert\overline{a}_i\rvert=m} \qbin{m}{\overline{a}_i}{\qm} \prod\limits_{k=1}^{3} \left( x_{ik}\otimes x_{kj} \right)^{a_{ik}},
\end{align*}
with the understanding that $\overline{a}_i = (a_{i1},a_{i2},a_{i3})$. The coaction \eqref{coaction} of a generic basis elements in \eqref{monomialbasis}, is then
\begin{align}\label{ca1}
 \Delta(z^{\vec{m}}) = \prod\limits_{i=1}^{3}\Delta(z_{i}^{m_i})= \sum\limits_{\substack{\rvert \overline{a}_1 \rvert = m_1\\\rvert \overline{a}_2 \rvert = m_2\\\rvert \overline{a}_3 \rvert = m_3}} \qbin{\vec{m}}{a}{\qm} \prod\limits_{i=1}^{3}\left( \prod\limits_{k=1}^{3} (x_{ik}\otimes x_{kj})^{a_{ik}} \right),
\end{align}
where all the products are to be expanded from left to right and from innermost to outermost. Then, observe that
\begin{align}\label{prodswap}
\prod\limits_{i=1}^{3} \left(\prod\limits_{k=1}^{3} x_{ik}^{a_{ik}} \right) &= x_{11}^{a_{11}} x_{12}^{a_{12}} x_{13}^{a_{13}} x_{21}^{a_{21}} x_{22}^{a_{22}} x_{23}^{a_{23}} x_{31}^{a_{31}} x_{32}^{a_{32}} x_{33}^{a_{33}}\nonumber\\
 &= x_{11}^{a_{11}} x_{21}^{a_{21}} x_{31}^{a_{31}} x_{12}^{a_{12}} x_{22}^{a_{22}} x_{32}^{a_{32}} x_{13}^{a_{13}} x_{23}^{a_{23}} x_{33}^{a_{33}} = \prod\limits_{k=1}^{3}\left( \prod\limits_{i=1}^{3} x_{ik}^{a_{ik}} \right).
\end{align}
Writing $n_k = \sum_{i=1}^{3} a_{ik}$ with $\rvert\vec{n}\rvert = \sum_{k=1}^3 n_k$, one finds with \eqref{Qa} that
\begin{align}\label{qaimplicit}
 \prod\limits_{i=1}^{3} \left(\prod\limits_{k=1}^{3} x_{kj}^{a_{ik}} \right) = Q(a) \prod\limits_{k=1}^{3} \left( x_{kj}^{a_{1k}+a_{2k}+a_{3k}} \right) = Q(a) z_{1}^{n_1}z_{2}^{n_2}z_{3}^{n_3} \equiv Q(a) z^{\vec{n}}.
\end{align}
Using \eqref{prodswap} and \eqref{qaimplicit} in \eqref{ca1} yields
\begin{align*}
 \Delta(z^{\vec{m}}) = \sum\limits_{\rvert\vec{n}\rvert = N} \sum\limits_{\substack{\,\rvert \overline{a}_j \rvert = m_j\\ \rvert \underline{a}_j \rvert = n_j}} \left[ Q(a) \qbin{\vec{m}}{a}{\qm} \prod\limits_{k=1}^{3}\left( \prod\limits_{i=1}^{3} x_{ik}^{a_{ik}} \right) \right] \otimes \prod\limits_{k=1}^{3} x_{kj}^{n_k},
\end{align*}
where we introduced $\underline{a}_j = (a_{1j}, a_{2j}, a_{3j})$ with the sums over $\rvert \overline{a}_j \rvert = m_j$ and $\rvert \underline{a}_j \rvert = n_j$ are sums over all the $\{ a_{ij} \}_{i,j=1,2,3}$ satisfying $\sum_{i=1}^{3} a_{ij} = n_j$ and $\sum_{j=1}^{3} a_{ij} = m_i$. From this expression, one directly identifies the matrix elements of $\mathcal{F}_q^{(N)}(\C^3)$.
\end{proof}
\end{proposition}

\newpage
\subsection{Unitarity}
 Unitary corepresentations can be constructed from the above corepresentations through normalization. We have:
 \begin{theorem}[Unitarity]\label{unitarity}
 The following $\qg$ corepresentation is unitary
 \begin{align*}
 \Delta: z^{\vec{m}} \qbin{N}{\vec{m}}{\qm}^{\frac{1}{2}} \longmapsto \sum\limits_{\rvert\vec{n}\rvert=N} t_{\vec{m},\vec{n}}^{(N)}\otimes z^{\vec{n}} \qbin{N}{\vec{n}}{\qm}^{\frac{1}{2}},
 \end{align*}
 with the matrix elements given by
 \begin{align}\label{mxLeft}
 t_{\vec{m},\vec{n}}^{(N)} = \sqrt{\qbin{N}{\vec{m}}{\qm} \qbin{N}{\vec{n}}{\qm}^{-1}} \sum\limits_{\substack{\,\rvert \overline{a}_j \rvert = m_j\\ \rvert \underline{a}_j \rvert = n_j}} Q(a) \qbin{\vec{m}}{a}{\qm} \prod\limits_{k=1}^{3}\left( \prod\limits_{i=1}^{3} x_{ik}^{a_{ik}} \right),
\end{align}
 and $Q(a)$ as in \eqref{Qa}.
 
 \begin{proof}
 A right $\qg$ comodule is constructed similarly as the left comodule constructed in \eqref{coaction}. Identifying the right comodule generators as $w_{j}=x_{ij}$ for $i$ fixed, one has now
\begin{align*}
 \Delta(w^{\vec{m}}) = \prod\limits_{j=1}^{3}\Delta(x_{ij}^{m_j})= \sum\limits_{\rvert \underline{b}_i \rvert = m_i} \qbin{\vec{m}}{b}{\qm} \prod\limits_{j=1}^{3}\left( \prod\limits_{k=1}^{3} (x_{ik}\otimes x_{kj})^{b_{kj}} \right),
\end{align*}
where the notation for the indices $b$ is the same as in Proposition \ref{mxelements}. Writing $n_k = b_{k1}+b_{k2}+b_{k3}$ such that
\begin{align*}
\prod\limits_{j=1}^{3} \left(\prod\limits_{k=1}^{3} x_{ik}^{b_{kj}} \right) = Q(b^{\top}) \prod\limits_{k=1}^{3}\left( \prod\limits_{j=1}^{3} x_{ik}^{b_{kj}} \right) = Q(b^{\top}) \prod\limits_{k=1}^{3} x_{ik}^{n_k} = Q(b^{\top}) w^{\vec{n}},
\end{align*}
to obtain, given that \eqref{Qa} is symmetric,
\begin{align}\label{unormalizedMxRight}
 \Delta(w^{\vec{m}})=\sum\limits_{\rvert\vec{n}\rvert=N} w^{\vec{n}} \otimes \tilde{h}_{\vec{n},\vec{m}}^{(N)}, \qquad \tilde{h}_{\vec{n},\vec{m}}^{(N)}= \sum\limits_{\substack{\rvert \overline{b}_i \rvert = n_i\\ \, \rvert \underline{b}_i \rvert = m_i}} Q(b) \qbin{\vec{m}}{b}{\qm} \prod\limits_{j=1}^{3} \left(\prod\limits_{k=1}^{3} x_{kj}^{b_{kj}}\right).
\end{align}
The generators of the left and right comodules are related by the $*$ operation in the following way
\begin{align*}
(w_{j})^{*} = (x_{ij})^{*} = S(x_{ji}) = S(z_{j}) \quad \implies \quad (w^{\vec{m}})^{*} = S(z^{\vec{m}}).
\end{align*}
Thus, knowing that $\Delta \circ S(x) = (S\otimes S)\circ \tau \circ \Delta$ for $\tau(x\otimes y) \equiv y\otimes x$, one has on the one hand
\begin{align}\label{deltastar1}
\Delta\, (w^{\vec{m}})^{*} = \Delta \circ S(z^{\vec{m}}) \longmapsto & \sum\limits_{\rvert\vec{n}\rvert=N} S(z^{\vec{n}}) \otimes S(h_{\vec{m},\vec{n}}) = \sum\limits_{\rvert\vec{n}\rvert=N} (w^{\vec{n}})^{*} \otimes S(h_{\vec{m},\vec{n}}). 
\end{align}
On the other hand, $\Delta$ being a $*$-homomorphism, one has
\begin{align}\label{deltastar2}
\Delta(w^{\vec{m}})^{*} \longmapsto \sum\limits_{\rvert\vec{n}\rvert=N} (w^{\vec{n}})^{*} \otimes (\tilde{h}_{\vec{n},\vec{m}})^{*}.
\end{align}
Knowing that the $w^{\vec{n}}$ are linearly independent, it follows from \eqref{deltastar1} and \eqref{deltastar2} that
\begin{align}\label{implicituni}
S\left(h_{\vec{m},\vec{n}}^{(N)}\right) = \left(\tilde{h}_{\vec{n},\vec{m}}^{(N)}\right)^{*}.
\end{align}
Comparing \eqref{unormalizedMxRight} with the matrix elements $h_{\vec{n},\vec{m}}^{(N)}$ of the left coaction \eqref{unormalizedMxLeft}, one can see that they only differ by the $q$-trinomial coefficient. However, it is easy to show that
\begin{align*}
 \qbin{N}{\vec{m}}{\qm} \qbin{\vec{m}}{a}{\qm} = \qbin{N}{\vec{n}}{\qm} \qbin{\vec{n}}{a}{\qm},
\end{align*}
so that with the proper normalization of the matrix elements, one has
\small
\begin{align*}
t_{\vec{n},\vec{m}}^{(N)} \equiv \sqrt{\qbin{N}{\vec{n}}{\qm}\qbin{N}{\vec{m}}{\qm}^{-1}} h_{\vec{n},\vec{m}}^{(N)} &= \sum\limits_{\substack{\rvert \overline{a}_j \rvert = n_j\\ \, \rvert \underline{a}_j \rvert = m_j}} Q(a) \sqrt{\qbin{\vec{n}}{a}{\qm}\qbin{\vec{m}}{a}{\qm}} \prod\limits_{k=1}^{3}\left( \prod\limits_{i=1}^{3} x_{ik}^{a_{ik}} \right)\\
&= \sqrt{\qbin{N}{\vec{m}}{\qm}\qbin{N}{\vec{n}}{\qm}^{-1}} \tilde{h}_{\vec{n},\vec{m}}^{(N)} \equiv \tilde{t}_{\vec{n},\vec{m}}^{(N)}.
\end{align*}
\normalsize
This normalization is equivalent to normalizing the basis elements of the left and right corepresentations as follows
\begin{align}\label{normbasis}
\underline{w}^{\vec{m}} \equiv \sqrt{\qbin{N}{\vec{m}}{\qm}} w^{\vec{m}}, \qquad \underline{z}^{\vec{m}} \equiv z^{\vec{m}} \,\sqrt{\qbin{N}{\vec{m}}{\qm}}.
\end{align}
With this normalization and \eqref{implicituni}, one has
\begin{align}\label{repunitarity}
S\left(t_{\vec{m},\vec{n}}^{(N)}\right) = \left(t_{\vec{n},\vec{m}}^{(N)}\right)^{*},
\end{align}
which establishes \cite[Ch. 4.1]{Chari1994} the unitarity of the corepresentations.
\end{proof}
 \end{theorem}
A direct corollary of Theorem \ref{unitarity} is the orthonormality of the matrix elements. Indeed, writing the $\qg$ product as $\nabla:\qg\otimes\qg\rightarrow\qg$, the hexagonal relation from the Hopf algebra structure gives
\begin{align}\label{hexagon}
 \sum\limits_{\rvert\vec{n}\rvert=N} t_{\vec{m},\vec{n}}\, S\left(t_{\vec{n},\vec{p}}\right) = \nabla \circ (1\otimes S) \circ \Delta\, \left(t_{\vec{m},\vec{p}}\right) = \eta \circ \epsilon\, \left( t_{\vec{m},\vec{p}} \right) = \delta_{\vec{m},\vec{p}},
\end{align}
where the last equality relies on the fact that the counit $\epsilon$ vanishes on off-diagonal generators of $\qg$ and also that the single term in \eqref{mxLeft} containing only diagonal generators has coefficient one. Upon using \eqref{repunitarity} in \eqref{hexagon}, one obtains
\begin{align}\label{orthonormality}
\sum\limits_{\rvert\vec{n}\rvert=N} t_{\vec{m},\vec{n}}\, t_{\vec{p},\vec{n}}^{*} &= \delta_{\vec{m},\vec{p}}, & \sum\limits_{\rvert\vec{m}\rvert=N} t_{\vec{m},\vec{n}}^{*}\, t_{\vec{m},\vec{p}} &= \delta_{\vec{n},\vec{p}},
\end{align}
where a similar argument is used to obtain the second identity.

\section{Matrix elements in $\qg$ representations}\label{mxinreps}
In this section, the matrix elements \eqref{mxLeft} are shown to be $q$-Krawtchouk polynomials. Following \cite{Koornwinder1989}, we first identify the matrix elements in the elementary representations $\pi_1$ and $\pi_2$ as univariate quantum $q$-Krawtchouk polynomials. Then, the matrix elements in the $\pi_{21}$ representation are identified as the Tratnik type bivariate quantum $q$-Krawtchouk polynomials. Finally, it is shown the matrix elements in the $\pi_{121}$ can be expressed in terms of the same polynomials.

\subsection{Elementary representations}
Consider the matrix elements \eqref{mxLeft} in the $\pi_1$ and $\pi_2$ representations. That these matrix elements are given in terms of univariate quantum $q$-Krawtchouk polynomials is a corollary of a previous result of Koornwinder \cite{Koornwinder1989}, as these $\qg$ repesentations where constructed from an $SU_q(2)$ representation. However, as we now work in an $\qg$ representation instead of the algebra itself a derivation in the current context is given. The univariate quantum $q$-Krawtchouk polynomials \cite{Koekoek1996} are defined\footnote{Part of the normalization of \cite{Koekoek1996} is here included in the polynomial $k_{n}(x;p,N,q)$ themselves.} by
\begin{align}\label{qKrawtchouk}
k_{n}(x;p,N,q) = (-1)^{n} \qpoc{q^{-N}}{q}{n} q^{n(n-1)/2} {}_2 \phi_{1} \pfq{q^{-n},q^{-x}}{q^{-N}}{q}{pq^{n+1}},
\end{align}
and are orthogonal with respect to the weight $w_{x}(p)^{2}$ for $x \in \{0,\cdots,N\}$, where
\begin{align*}
 w_{x}(p) = \left[ (-1)^{N-x} q^{x(x-1)/2}\qbin{N}{x}{q} \frac{\qpoc{pq}{q}{N-x}}{\qpoc{q}{q}{N}} p^{-N} q^{-N(N+1)/2} \right]^{1/2},
\end{align*}
with normalization
\begin{align*}
 \Theta_n(p) = \frac{q^{-n(n-1)/2}}{\qpoc{q^{-N}}{q}{n}} \left[ (-1)^n q^{n(n+1)/2-Nn} \qbin{N}{n}{q} \frac{\qpoc{q}{q}{N}}{\qpoc{qp}{q}{n}}  \right]^{1/2}.
\end{align*}
One has the following proposition.
\begin{proposition}\label{uniMX}
The matrix elements \eqref{mxLeft} evaluated in the elementary representations $\pi_i$ with $i=1,2$ as defined in \eqref{pi1rep} and \eqref{pi2rep}, are shift operators given by
\begin{align}
 \pi_i\left(t_{\vec{m},\vec{n}}^{(N)}\right)\ket{k} = (\delta_{i,2}\delta_{m_1,n_1}+\delta_{i,1}\delta_{N-m_1-m_2,N-n_1-n_2}) t_{m_i,n_i,T_i}(k) \ket{k+T_i-m_i-n_i},\label{unishift}\\
 t_{m,n,T}(k-T+n) \equiv (-1)^{n-m} w_{n}(q^{-2(k+1)}) \Theta_{m}(q^{-2(k+1)}) k_m(q^{-2n};q^{-2(k+1)},T,\qp).\label{uniscalar}
\end{align}
\begin{proof}
 Looking at \eqref{pi1rep}, the monomials in the $\qg$ generators in \eqref{mxLeft} evaluate to
\begin{multline*}
 \pi_{1} \left( \prod\limits_{i,k=1}^{3} x_{ik}^{a_{ik}} \right) \ket{k} =\delta_{a_{33},m_{3}} \prod\limits_{i=1}^{2} \delta_{a_{3i},0}\,\delta_{a_{i3},0}\\
 \times (-1)^{a_{21}} q^{a_{12}(k+a_{22}+1) + a_{21}(k+a_{22})} \sqrt{\qpoc{q^{2k+2}}{\qp}{a_{22}}\qpoc{q^{2(k+a_{22})}}{\qm}{a_{11}}}\ket{k+a_{22}-a_{11}}.
\end{multline*}
Using the above and writing $T=N-m_3=N-n_3$, the matrix elements simplifies to 
\begin{multline*}
 \pi_1\left(t_{\vec{m},\vec{n}}^{(N)}\right)\ket{k} = \sum\limits_{a_{21}=0}^{m_2} \qbin{m_1}{a_{11}}{\qm} \qbin{m_2}{a_{21}}{\qm} q^{-a_{21}(a_{12})}(-1)^{a_{21}}q^{a_{12}(k+a_{22}+1) + a_{21}(k+a_{22})}\\
 \times\sqrt{\qbin{N}{\vec{m}}{\qm}\qbin{N}{\vec{n}}{\qm}^{-1}}\sqrt{\qpoc{q^{2k+2}}{\qp}{a_{22}}\qpoc{q^{2(k+a_{22})}}{\qm}{a_{11}}}\ket{k+a_{22}-a_{11}}.
\end{multline*}
Taking the following parametrization of the summation indices
\begin{align*}
 \begin{pmatrix}
  a_{11} & a_{12}\\
  a_{21} & a_{22}
 \end{pmatrix}
  =
 \begin{pmatrix}
  T-n_2-i & n_2-m_2+i\\
  i & m_2-i
 \end{pmatrix},
\end{align*}
one has
\begin{multline*}
 \pi_1\left(t_{\vec{m},\vec{n}}^{(N)}\right)\ket{k} = (-1)^{T-n_2} q^{(n_2-m_2)(k+m_2+1)+n_1(n_1+1)}\\
 \times \sqrt{\qpoc{q^{2(k+1)}}{\qp}{m_2}\qpoc{q^{2(k+m_2-n_1+1)}}{\qp}{n_1}} \frac{\qpoc{q^{-2m_1}}{\qp}{n_1}}{\qpoc{\qp}{\qp}{n_1}} \ket{k-n_1+m_2}\\
 \times \sqrt{\qbin{N}{\vec{m}}{\qm}\qbin{N}{\vec{n}}{\qm}^{-1}} \sum\limits_{i=0}^{m_2} \frac{\qpoc{q^{-2n_1}}{\qp}{i}\qpoc{q^{-2m_2}}{\qp}{i}}{\qpoc{q^{2(n_2-m_2+1)}}{\qp}{i}\qpoc{q^{-2(k+m_2)}}{\qp}{i}\qpoc{\qp}{\qp}{i}}q^{2i}.
\end{multline*}
Using Jackson's identity \cite[(III.5)]{Gasper2004},
% $$ \qpoc{q^{-n} \frac{z}{b}}{q}{n} {}_3\phi_2 \pfq{q^{-n},b,0}{qb/z,c}{q}{q} = {}_2\phi_1 \pfq{q^{-n},c/b}{c}{q}{z},$$
one can write the ${}_3\phi_2$ as a ${}_2\phi_1$.
% \begin{multline*}
%  \pi_1\left(t_{\vec{m},\vec{n}}^{(N)}\right)\ket{k} = (-1)^{T-n_2} q^{(n_2-m_2)(k+m_2+1)+n_1(n_1+1)}\\
%  \times \sqrt{\frac{\qpoc{q^{2(k+m_2-n_1+1)}}{\qp}{n_1}}{\qpoc{q^{2(k+1)}}{\qp}{m_2}}} \frac{\qpoc{q^{-2m_1}}{\qp}{n_1}}{\qpoc{\qp}{\qp}{n_1}} \ket{k-n_1+m_2}\\
%  \times \sqrt{\qbin{N}{\vec{m}}{\qm}\qbin{N}{\vec{n}}{\qm}^{-1}} {}_2\phi_1 \pfq{q^{-2m_2},q^{2(n_1+n_2-m_2+1)}}{q^{2(n_2-m_2+1)}}{\qp}{q^{2(k+1-n_1+m_2)}}.\\
% \end{multline*}
% Finally one
Then, reversing the order of summation using \cite[exer. 1.4 (ii)]{Gasper2004}
% \begin{align*}
%  {}_2\phi_1 \pfq{q^{-n},b}{c}{q}{z} = (-1)^n q^{-n(n+1)/2} \frac{\qpoc{b}{q}{n}}{\qpoc{c}{q}{n}} z^n {}_2\phi_1 \pfq{q^{-n},q^{-n+1} c^{-1} }{q^{-n+1}b^{-1}}{q}{\frac{q^{n+1} c}{b z}},
% \end{align*}
, after shifting the parameter $k$ by $-n_2$ and using $q$-Pochhammer symbol identities, leads to
\begin{multline*}
 \pi_1\left(t_{\vec{m},\vec{n}}^{(N)}\right)\ket{k} = (-1)^{m_2} q^{k(n_2+m_2)+n_2(m_2+1)-2n_1n_2} \ket{k-n_1+m_2}\\
 \times \qbin{T}{n_{1}}{\qp} \sqrt{\qbin{N}{\vec{m}}{\qm}\qbin{N}{\vec{n}}{\qm}^{-1}\frac{\qpoc{q^{2(k-n_1+1)}}{\qp}{n_1}}{\qpoc{q^{2(k-n_1+1)}}{\qp}{m_2}}} {}_2\phi_1 \pfq{q^{-2m_2},q^{-2n_2}}{q^{-2(n_1+n_2)}}{\qp}{q^{-2k}}.
\end{multline*}
We now use one of Heine's transformation formulas \cite[(III.3)]{Gasper2004}
% \begin{align*}
% {}_2\phi_1 \pfq{q^{-2m},q^{-2p}}{q^{-2N}}{\qp}{q^{-2k+2p}} = \frac{1}{\qpoc{q^{-2k+2p}}{\qp}{N-m-p}} {}_2\phi_1 \pfq{q^{-2(N-m)},q^{-2(N-p)}}{q^{-2N}}{\qp}{q^{-2k+2(N-m)}},
% \end{align*}
to obtain
\begin{multline}\label{unipi1mx}
 \pi_1\left(t_{\vec{m},\vec{n}}^{(N)}\right)\ket{k-T+n_1} = \delta_{N-m_1-m_2,N-n_1-n_2}(-1)^{n_1} q^{n_1^2-n_1T} \\
 \times \Bigg[(-1)^{T-n_1+m_1} q^{2T(k+1)-T(T+1)-m_1(m_1-1)+n_1(n_1-1)} \frac{\qpoc{q^{-2k}}{\qp}{T-n_1}}{\qpoc{q^{-2k}}{\qp}{m_1}}\Bigg]^{1/2}\\
 \times \qbin{T}{n_{1}}{\qp} \sqrt{\qbin{N}{\vec{m}}{\qm}\qbin{N}{\vec{n}}{\qm}^{-1}} {}_2\phi_1 \pfq{q^{-2m_1},q^{-2n_1}}{q^{-2T}}{\qp}{q^{-2k+2m_1}}\ket{k-m_1}.
\end{multline}
Proceeding similarly, one obtains for $\pi_{2}$
\begin{multline}\label{unipi2mx}
 \pi_2\left(t_{\vec{m},\vec{n}}^{(N)}\right)\ket{k-T+n_2} = \delta_{m_{1},n_{1}}(-1)^{n_2} q^{n_2^2-n_2T} \qbin{T}{n_{2}}{\qp} \sqrt{\qbin{N}{\vec{m}}{\qm}\qbin{N}{\vec{n}}{\qm}^{-1}} \\
 \times \Bigg[(-1)^{T-n_2+m_2} q^{2T(k+1)-T(T+1)-m_2(m_2-1)+n_2(n_2-1)} \frac{\qpoc{q^{-2k}}{\qp}{T-n_2}}{\qpoc{q^{-2k}}{\qp}{m_2}}\Bigg]^{1/2}\\
 \times {}_2\phi_1 \pfq{q^{-2m_2},q^{-2n_2}}{q^{-2T}}{\qp}{q^{-2k+2m_2}}\ket{k-m_2}.
\end{multline}
From \eqref{unipi1mx} and \eqref{unipi2mx}, one directly finds \eqref{unishift} and \eqref{uniscalar}, which concludes the proof. 
\end{proof}
\end{proposition}

\subsection{Product representations}
We now demonstrate that the matrix elements in the representation $\pi_{21}$ are espressed in terms of bivariate quantum $q$-Krawtchouk polynomials of the Tratnik type. These polynomials are defined in \cite{Gasper2007} as
\begin{align}\label{qTratnik}
 K_{n,m}(x,y;u,v,N,q) = k_{n}(x;v^{-2},x+y,q) k_{m}(x+y-n;u^{-2},N-n,q),
\end{align}
with $k_n(x;p,N,q)$ as in \eqref{qKrawtchouk}. They are orthogonal with respect to the weight $W_{n_{1},n_{2}}^{(N)}(u,v)^2$ where
\begin{multline}\label{weightTratnik}
 W_{n_{1},n_{2}}^{(N)}(u,v) = \Bigg[(-1)^{N-n_{1}}q^{2v(n_{1}+n_{2})} q^{n_1(n_1-1)}\\
 \times \qbin{N}{\vec{n}}{\qp} \qpoc{q^{-2v}}{\qp}{n_{2}} \qpoc{q^{-2u}}{\qp}{N-n_{1}-n_{2}} q^{2N(u+1)} q^{-N(N+1)}\Bigg]^{1/2},
\end{multline}
with the normalization given by
\begin{multline}\label{normTratnik}
N_{m_{1},m_{2}}^{(N)}(u,v) = q^{-m_{1}(m_{1}-1)-m_{2}(m_{2}-1)} q^{-m_{1}(u+1)} \frac{ \qpoc{\qp}{\qp}{N-m_{1}-m_{2}} }{\qpoc{\qp}{\qp}{N}}\\
\times \left[ (-1)^{m_1+m_2} \qbin{N}{\vec{m}}{\qp} \frac{q^{2N(m_{1}+m_{2})+4m_{1}+2m_{2}-2m_{1}m_{2}-m_1(m_1-1)-m_2(m_2-1)} }{ \qpoc{q^{-2u}}{\qp}{m_2} \qpoc{q^{-2v}}{\qp}{m_1} } \right]^{1/2}.
\end{multline}
With these notations, one has the following proposition.
\begin{proposition}
The matrix elements \eqref{mxLeft} evaluated in $\pi_{21}$ as defined in \eqref{pi12} are shift operators specified by
\begin{align}\label{bishift}
 \pi_{21}\left(t_{\vec{m},\vec{n}}^{(N)}\right)\ket{u,v} = t_{\vec{m},\vec{n}}^{(21)}(u,v)\ket{u-m_2+N-n_1-n_2,v+n_2-m_1},
\end{align}
where
\begin{multline}\label{tratnikscalar}
 t_{\vec{m},\vec{n}}^{(21)}(u-N+n_1+n_2,v-n_2) =(-1)^{m_{1} - n_{2}}\\
 \times W_{n_{1},n_{2}}^{(N)}(u,v) N_{m_{1},m_{2}}^{(N)}(u,v) K_{m_1,m_2}(n_1,n_2;q^{u+1},q^{v+1},N,\qp),
\end{multline}
are the normalized and weighted bivariate quantum $q$-Krawtchouk polynomials of the Tratnik type. 
\begin{proof}
From \eqref{pi12}, one has
\begin{align}\label{12action}
\pi_{21}\left(t_{\vec{m},\vec{n}}^{(N)}\right)\ket{u,v} = (\pi_{2} \otimes \pi_{1}) \circ \Delta\left(t_{\vec{m},\vec{n}}^{(N)}\right)\ket{u,v} = \sum\limits_{\rvert\vec{k}\rvert=N} \pi_{2}\left(t_{\vec{m},\vec{k}}^{(N)}\right)\ket{u} \otimes \pi_{1}\left(t_{\vec{k},\vec{n}}^{(N)}\right)\ket{v}.
\end{align}
Using \eqref{unishift} of Proposition \ref{uniMX} in \eqref{12action}, one sees the Kronecker deltas $\delta_{m_1,k_1}$ and $\delta_{k_1+k_2,n_1+n_2}$ remove the sums and one has
\begin{align*}
 k_1 = m_1, \quad k_2 = (n_1+n_2-m_1),
\end{align*}
which implies $T_1 = n_1+n_2$ and $T_2 = N-m_1$. One can thus identify \eqref{12action} as the shift operator in \eqref{bishift}. Then, shifting $u$ by $n_1+n_2-N$ and $v$ by $n_2$ in \eqref{12action} one has that
\begin{multline}\label{scalar1}
 t_{\vec{m},\vec{n}}^{(21)}(u-N+n_1+n_2,v-n_2)=\\t_{m_2,n_1+n_2-m_1,N-m_1}(u-N+n_1+n_2)t_{m_1,n_1,n_1+n_2}(v-n_2).
\end{multline}
Using \eqref{uniscalar} from Proposition \ref{uniMX} in \eqref{scalar1}, one obtains by involved but direct computations
% \begin{multline}\label{scalar1}
%  t_{m_2,n_1+n_2-m_1,N-m_1}(u-N+n_1+n_2) = (-1)^{n_1+n_2-m_1-m_2}q^{(n_1+n_2-m_1)^2-(n_1+n_2-m_1)(N-m_1)}\\
%  \times \qbin{N-m_1}{n_1+n_2-m_1}{\qp} \Big[ q^{(n_1+n_2-m_1)(n_1+n_2-m_1-1)+2(N-m_1)(u+1)-(N-m_1)(N-m_1+1)-3m_2(m_2-1)} \Big]^{-1/2}\\
%  \times \sqrt{(-1)^{N-n_1-n_2+m_2}\qbin{N}{\vec{m}}{\qm}\qbin{N}{\vec{k}}{\qm}^{-1} \frac{\qpoc{q^{-2u}}{\qp}{N-n_1-n_2}}{\qpoc{q^{-2u}}{\qp}{m_2}} }\, \frac{k_{m_2}(n_1+n_2-m_1;q^{-2(u+1)},N-m_1,\qp)}{\qpoc{q^{-2(N-m_1)}}{\qp}{m_2}},
% \end{multline}
% and
% \begin{multline}\label{scalar2}
%  t_{m_1,n_1,n_1+n_2}(v-n_2) = (-1)^{n_1-m_1}q^{n_1^2-n_1(n_1+n_2)}\\
%  \times \qbin{n_1+n_2}{n_1}{\qp} \Big[ q^{2(n_1+n_2)(v+1)-(n_1+n_2)(n_1+n_2+1)-3m_1(m_1-1)+n_1(n_1-1)} \Big]^{1/2}\\
%  \times \sqrt{(-1)^{m_1+n_2} \qbin{N}{\vec{k}}{\qm}\qbin{N}{\vec{n}}{\qm}^{-1} \frac{\qpoc{q^{-2v}}{\qp}{n_2}}{\qpoc{q^{-2v}}{\qp}{m_1}} }\, \frac{k_{m_1}(n_1;q^{-2(v+1)},n_1+n_2,\qp)}{\qpoc{q^{-2(n_1+n_2)}}{\qp}{m_1}}.
% \end{multline}
% The scalar factor of the shift operator is the product of \eqref{scalar1} and \eqref{scalar2}. By involved but direct computations one obtains
\begin{multline*}
 t_{\vec{m},\vec{n}}^{(21)}(u-N+n_1+n_2,v-n_2) = \\
 (-1)^{m_{1} - n_{2}}\,  W_{n_{1},n_{2}}^{(N)}(u,v) N_{m_{1},m_{2}}^{(N)}(u,v) K_{m_1,m_2}(n_1,n_2;q^{u+1},q^{v+1},N,\qp),
\end{multline*}
which concludes the proof.
\end{proof}
\end{proposition}
The expression in \eqref{tratnikscalar} for the scalar factor of the matrix elements \eqref{mxLeft} evaluated in $\pi_{21}$ has been obtained in a related but different approach in \cite{Genest2013}, see also \cite{Groenevelt2018}. There, \eqref{tratnikscalar} arises as an expression for the matrix elements in symmetric representations of unitary elements of $U_q(\mathfrak{su}(3))$ constructed from $q$-exponentials. This correspondance is expected in view of the duality \cite{Burroughs1990}, \cite[Ch. 7]{Chari1994} between quantum algebras and groups and has been discussed \cite{Floreanini1993a} in the context of $q$-special functions. The parameters of the polynomials in \eqref{tratnikscalar} is discrete. However, one recovers the usual $q$-Krawtchouk polynomials with continuous parameters by extension with analytic continuation.

\subsection{Representation corresponding to the longest Weyl group element}
We now study the matrix elements in $\pi_{121}$. In this case, one has
\begin{align*}
 \pi_{121}\left( t_{\vec{m},\vec{n}}^{(N)} \right)\ket{t,u,v} &= (\pi_1 \otimes \pi_{21})\circ \Delta\left( t_{\vec{m},\vec{n}}^{(N)} \right) \ket{t,u,v}\\
 &= \sum\limits_{\rvert k \rvert = N} \pi_1\left( t_{\vec{m},\vec{k}}^{(N)} \right)\ket{t} \otimes \pi_{21}\left(t_{\vec{k},\vec{n}}^{(N)}\right)\ket{u,v}.
\end{align*}
Using \eqref{unishift}, \eqref{uniscalar}, \eqref{bishift} and \eqref{tratnikscalar} in the above, one obtains
\begin{multline}\label{pi121shift}
 \pi_{121}\left( t_{\vec{m},\vec{n}}^{(N)} \right)\ket{t,u,v} = \sum\limits_{k_1=0}^{m_1+m_2} (-1)^{k_1-m_1} t_{m_1,k_1,T_1}(t) t_{\vec{k},\vec{n}}^{(21)}(u,v)\\
 \times\ket{t+m_2-k_1,u+N-m_1-m_2-n_1-n_2+k_1,v+n_2-k_1}.
\end{multline}
As $t_{m_1,k_1,T_1}(t)$ does not depend on the variables, one can identify the scalar coefficients of \eqref{pi121shift} as bivariate $q$-Krawtchouk polynomials normalized by a factor expressed as a univariate $q$-Krawtchouk polynomials. Thus, studying the matrix elements in $\pi_{121}$ leads to the same polynomials. Indeed, the unitarity of the corepresentations is equivalent to the orthogonality relation of the Tratnik bivariate $q$-Krawtchouk polynomials \eqref{qTratnik}:
\begin{align*}
 \delta_{\vec{n},\vec{p}} \bra{a',b',c'}a,b,c\rangle = \sum\limits_{\rvert\vec{m}\rvert=N} \bra{a',b',c'}\pi_{121}\left( t_{\vec{m},\vec{n}}^{(N)} \right)^{*} \pi_{121}\left( t_{\vec{m},\vec{p}}^{(N)} \right)\ket{a,b,c},
\end{align*}
which, after shifting $b$ by $p_1+p_2$ and $c$ by $-p_2$, is given by
\begin{multline*}
 \delta_{\vec{n},\vec{p}} \bra{a',b',c'}a,b+p_1+p_2,c-p_2\rangle =\\
 \sum\limits_{\rvert\vec{m}\rvert=N} \sum\limits_{k_1=0}^{m_1+m_2} \sum\limits_{l_1=0}^{m_1+m_2} (-1)^{k_1+l_1} t_{m_1,k_1,T_1}(a) t_{m_1,l_1,T_1}(a-k_1+l_1)\\
 \times t_{\vec{k},\vec{p}}^{(21)}(b+p_1+p_2,c-p_2) t_{\vec{l},\vec{n}}^{(21)}(b+n_1+n_2+k_1-l_1,c-n_2-k_1+l_1)\\
 \times \bra{a',b',c'}a-k_1+l_1,b+n_1+n_2+k_1-l_1,c-n_2-k_1+l_1\rangle.
\end{multline*}
Noticing that the overlap $\bra{a'}a-k_1+l_1\rangle$ fixes the difference $l_1-k_1 \equiv s$, one can rearrange the sums to obtain
\begin{multline*}
 \delta_{\vec{n},\vec{p}} \bra{a-s,b',c'}a,b+p_1+p_2,c-p_2\rangle =\\
 \sum\limits_{\rvert\vec{k}\rvert=N} \sum\limits_{m_1=0}^{k_1+k_2} t_{\vec{k},\vec{p}}^{(21)}(b+p_1+p_2,c-p_2) t_{\vec{k}+s(\vec{e}_1-\vec{e}_2),\vec{n}}^{(21)}(b+n_1+n_2-s,c-n_2+s)\\
 \times (-1)^s t_{m_1,k_1,T_1}(a) t_{m_1,k_1-s,T_1}(a-s) \braket{b',c'}{b+n_1+n_2-s,c-n_2+s}.
\end{multline*}
Using the univariate $q$-Krawtchouk dual orthogonality relation \cite{Koekoek1996} in the sum over $m_1$ forces $s$ to vanish so that one has
\begin{multline*}
 \delta_{\vec{n},\vec{p}} \braket{b',c'}{b+p_1+p_2,c-p_2} =\\
 \sum\limits_{\rvert\vec{k}\rvert=N} t_{\vec{k},\vec{p}}^{(21)}(b+p_1+p_2,c-p_2) t_{\vec{k},\vec{n}}^{(21)}(b+n_1+n_2,c-n_2)\braket{b',c'}{b+n_1+n_2,c-n_2},
\end{multline*}
which one can recognize as the dual orthogonality relation of the bivariate $q$-Krawtchouk polynomials of the Tratnik type.

\section{Reducible tensor products}\label{redrep}
The algebraic interpretation of the multivariate $q$-Krawtchouk polynomials presented in this paper can be used to derive identities for these polynomials. Consider for example the reducible tensor product of the $SU_{q}(2)$ representations $\tau_\alpha\otimes\tau_\beta$ with $\alpha,\beta \in S^1$. It is known \cite{Koornwinder1991} to decompose into the direct integral
\begin{align*}
\rho = \int\limits_{S^1}^{\oplus} \tau_\gamma d\gamma,
\end{align*}
 with the intertwiner $\Lambda:\tau_\alpha \otimes \tau_\beta \rightarrow \rho$ acting as follows on the basis vectors \cite{Groenevelt2014}
\begin{align*}
\Lambda\,\ket{v,t} = \sum\limits_{w\in \mathbb{N}} \alpha^{t-w}\beta^{w-v}\, \bar{p}_v(q^{2w};q^{2\rvert t-v \rvert};\qp)\,\frac{\gamma^{-v}}{\sqrt{2\pi}} \otimes \ket{w},
\end{align*}
where the Clebsch-Gordan coefficients, given by
\begin{multline*}
 \bar{p}_v(q^{2w};q^{2\rvert t-v \rvert};\qp) = (-1)^{v+w} \sqrt{\frac{q^{2(w-v)(\rvert t-v\rvert+1)} \qpoc{q^{2\rvert t-v\rvert +2}}{\qp}{\infty} \qpoc{q^{2\rvert t-v \rvert +2}}{\qp}{v} }{\qpoc{\qp}{\qp}{v}\qpoc{\qp}{\qp}{w} } }\\
 \times {}_2\phi_{1}\pfq{q^{-2v},0}{q^{2\rvert t-v \rvert+2}}{\qp}{q^{2w+2}},
\end{multline*}
are the weighted and normalized Wall polynomials \cite{Koekoek1996}. This result can be used in two ways to calculate the matrix elements in the $\pi_{211}$ representation. One has, on the one hand,
\begin{multline*}
\pi_{211}\left( t_{\vec{m},\vec{n}}^{(N)} \right) \ket{u,v,t} = \int\limits_{S^1}^{\oplus} d\kappa \sum\limits_{k_1=0}^{N} \sum\limits_{w' \in \mathbb{N}} (\sqrt{2\pi})^{-1/2} \ket{u-m_{2}+n_{3},w'}\\
 \times \alpha^{t+n_2-k_1-w'}\beta^{w'-v-n_1-n_2+k_1+m_1} \kappa^{m_1+k_1-n_1-n_2-v} t_{\vec{m},(k_1,n_1+n_2-k_1)}^{(21)}(u,v) t_{k_{1},n_{1},T_1}(t)\\
  \times \bar{p}_{v+n_1+n_2-k_1-m_1}(q^{2w'};q^{2\rvert t+m_1-v-n_1 \rvert};\qp),
\end{multline*}
 and, on the other hand,
 \begin{multline*}
 \pi_{211}\left( t_{\vec{m},\vec{n}}^{(N)} \right) \ket{u,v,t} =\\
 \int\limits_{S^1}^{\oplus} d\gamma\sum\limits_{w \in \mathbb{N}}\frac{\alpha^{t-w}\beta^{w-v}\gamma^{-v}}{\sqrt{2\pi}}\, \bar{p}_v(q^{2w};q^{2\rvert t-v \rvert};\qp)\, \pi_{21}\left(t_{\vec{m},\vec{n}}^{(N)}\right)\ket{u,w} = \\
 \int\limits_{S^1}^{\oplus} d\gamma\sum\limits_{w \in \mathbb{N}}\frac{\alpha^{t-w}\beta^{w-v}\gamma^{-v}}{\sqrt{2\pi}}\, \bar{p}_v(q^{2w};q^{2\rvert t-v \rvert};\qp)\, t_{\vec{m},\vec{n}}^{(21)}(u,w)\ket{u-m_{2}+n_{3},w+n_{2}-m_1}.
\end{multline*}
Taking the inner product on $S^{1} \otimes V_\alpha \otimes V_\gamma$ of both expression with
$
\frac{\gamma^{-v}}{\sqrt{2\pi}}\otimes\ket{u-m_2+n_3,w+n_2-m_1},
$
for a fixed but arbitrary $w$, one gets
\begin{multline*}
 \bar{p}_v(q^{2w};q^{2\rvert t-v \rvert};\qp)\, t_{\vec{m},\vec{n}}^{(21)}(u,w)= \sum\limits_{k_1=0}^{N} \alpha^{m_1-k_1}\beta^{k_1-n_1} t_{k_{1},n_{1},n_1+n_2}(t)\\
 \times \bar{p}_{v+n_1+n_2-k_1-m_1}(q^{2(w+n_2-m_1)};q^{2\rvert t+m_1-v-n_1 \rvert};\qp) t_{\vec{m},(k_1,n_1+n_2-k_1)}^{(21)}(u,v).
\end{multline*}
 Upon taking $\alpha=\beta=\gamma=-1$ while shifting $u$ by $n_1+n_2-N$, $v$ by $-n_1-n_2$, $w$ by $-n_2$ and $t$ by $-n_2$, one can use \eqref{uniscalar} and \eqref{tratnikscalar} to express the above in terms of the polynomials, obtaining
\begin{multline}\label{identity}
 \bar{p}_{v-n_1-n_2}(q^{2w-2n_2};q^{2\rvert t+n_1-v \rvert};\qp)\, K_{m_1,m_2}(n_1,n_2;q^{u+1},q^{w+1},N,\qp)=\\
 \sum\limits_{j=0}^{N} C_{\vec{m},\vec{n},j}^{(N)}(u,v,t)\, \bar{p}_{v-j-m_1}(q^{2(w-n_2-m_1)};q^{2\rvert t+m_1-v \rvert};\qp)\\
 \times k_j(q^{-2n_1};q^{-2(t+1)},n_1+n_2,\qp) K_{m_1,m_2}(j,n_1+n_2-j;q^{u+1},q^{v-j+1},N,\qp),
\end{multline}
where
\begin{multline*}
 C_{\vec{m},\vec{n},j}^{(N)}(u,v,t) =\\ (-1)^{m_1-n_1}\, w_{n_1}(q^{-2(t+1)}) \Theta_{j}(q^{-2(t+1)}) \frac{W_{j,n_1+n_2-j}^{(N)}(u,v-j)}{W_{n_{1},n_{2}}^{(N)}(u,w)} \sqrt{\frac{\qpoc{q^{-2w}}{\qp}{m_1}}{\qpoc{q^{-2v+2j}}{\qp}{m_1}}}.
\end{multline*}
It follows that \eqref{identity} provides an identity for the product of a bivariate quantum $q$-Krawtchouk polynomial and a Wall polynomial.

\section*{Conclusion}
This paper identified the matrix elements of the $\qg$ symmetric corepresentations as the bivariate quantum $q$-Krawtchouk polynomials of the Tratnik type. This was done by first constructing the symmetric unitary corepresentations and obtaining abstract expressions for the matrix elements and then evaluating these expressions in $\qg$ representations. These results thus provide an algebraic interpretation for the bivariate $q$-Krawtchouk polynomials within the quantum group setting. Moreover, this identification of the matrix elements evaluated in $\qg$ representations is complete in the sense that it held for the most general irreducible $*$-representation. Finally, this paper illustrated how the quantum group interpretation could be exploited to obtain properties of the $q$-Krawtchouk polynomials.

The results presented here should admit a generalization to the generic case of $SU_q(N)$, that would yield a similar algebraic interpretation of the multivariate quantum $q$-Krawtchouk polynomials of the Tratnik type with valuable outcomes. In view of the relation between reducibility of $SU_q(3)$ representations and an identity for orthogonal polynomials, this algebraic interpretation might prove useful to obtain identities for the polynomials.

\section*{Acknowledgements}
G.B. would like to thank the Institute for Mathematics, Astrophysics and Particle Physics at Radboud Universiteit for its hospitality and Paul Terwilliger for stimulating discussions. E.K. thanks the Centre de Recherches Math\'ematiques at the Universit\'e de Montr\'eal for its hospitality. G.B. holds a scholarship of the Natural Science and Engineering Research Council of Canada (NSERC). The research of L.V. is supported by a Discovery Grant from NSERC.

\bibliographystyle{abbrv}
\bibliography{q-krawtchouk.bib}

\end{document}